\newcommand{\SMY}{{M^2_{Y^{++}}}}
\newcommand{\SMX}{{M^2_{Y^{+}}}}
\newcommand{\MY}{{M_{Y^{++}}}}
\newcommand{\MX}{{M_{Y^{+}}}}

\newcommand{\suu}{$\mathrm{SU_c(3)} \times
\mathrm{SU_L(3)} \times \mathrm{U_X(1)}$}
\newcommand{\sm}{$\mathrm{SU_c(3)} \times
\mathrm{SU_L(2)} \times \mathrm{U_Y(1)}$~}

\documentstyle[aps,prd,epsfig,amsmath]{revtex}

\begin{document}

\tightenlines
\title{Bilepton gauge boson contribution to the static electromagnetic properties
of the W boson in the minimal 331 model}
\author{G. Tavares-Velasco}
\address{Departamento de F\'\i sica, CINVESTAV, Apartado Postal 14-740,
07000, M\' exico, D. F., M\' exico}
\author{J. J. Toscano}
\address{Facultad de Ciencias F\'\i sico Matem\' aticas, Benem\' erita
Universidad Aut\' onoma de Puebla, Apartado Postal 1152, 72000, Puebla,
Pue., M\' exico}

\date{\today}
\maketitle
\begin{abstract}
We present a complete calculation of the singly and doubly charged
gauge bosons (bileptons) contribution to the static properties of
the $W$ boson in the framework of the minimal 331 model, which
accommodates the bileptons in an $\mathrm{SU_L(2)}$ doublet. A
nonlinear $R_\xi$ gauge is used and a slightly modified version of
the Passarino-Veltman reduction scheme is employed as in this case
the Gram determinant vanishes. It is found that the bilepton
contribution is of the same order of magnitude as those arising
from other weakly coupled renormalizable theories, like the
two-Higgs doublet model and supersymmetry. The heavy mass limit is
explored and the nontrivial decoupling properties of bileptons are
discussed. Although there is a close resemblance with the
contribution of an $\mathrm{SU_L(2)}$ fermion doublet, in the case
of the bilepton doublet the decoupling theorem does remain valid.
As a by-product, we present a detailed study of the trilinear and
the quartic vertices involving the bileptons and the standard
model gauge bosons.
\end{abstract}
\draft

\section{Introduction}
\label{introduction}

One of the main goals of the next generation of high energy
colliders will be to probe the Yang-Mills sector of the standard
model (SM). In order to measure the $WW\gamma$ and $WWZ$ gauge
couplings, the most promising production modes are $WZ$ and
$W\gamma$ at hadron colliders and $WW$ at $e^+e^-$ linear
colliders. With a high experimental accuracy, these modes would
allow to test these couplings beyond tree level, which is
essential for studying the gauge cancellations that arise at
one-loop level. The study of these couplings also offers a unique
opportunity to find any evidence of heavy physics lying beyond the
Fermi scale. In particular, the on-shell electromagnetic
properties of the $W$ boson have been the subject of constant
interest since they can be sensitive to new physics effects. These
quantities are the anomalous (one-loop) magnetic dipole moment and
the electric quadrupole moment, which are characterized by two
parameters denoted by $\Delta \kappa$ and $\Delta Q$. They appear
as coefficients of Lorentz structures of canonical dimension 4 and
6, respectively. In the SM, both $\Delta \kappa$ and $\Delta Q$
vanish at tree level. This means that these parameters can only
receive contributions at one-loop level in any renormalizable
theory and may be sensitive to new physics effects, which might
compete with the SM contribution. We will see below that a high
precision measurement of $\Delta Q$ can only be useful to looking
for physics effects not very far beyond the Fermi scale. In
contrast, $\Delta \kappa$ may be sensitive to heavy physics
effects. Within the SM, the one-loop contributions to $\Delta Q$
and $\Delta \kappa$ from the gauge bosons, the Higgs scalar and
massless fermions were studied in \cite{Bardeen}, whereas the top
quark effects were analyzed later \cite{Couture1}. The sensitivity
of these quantities to new physics effects has also been studied
within some specific models, like the two-Higgs doublet model
\cite{Couture2} and supersymmetric theories \cite{SUSY}. Further
studies were also done within models with an extra ${Z^\prime}$
boson \cite{Sharma}, composite particles \cite{Rizzo}, and an
extra $W$ boson \cite{Larios}. Both $\Delta Q$ and $\Delta \kappa$
have also been parametrized in a model independent way by using an
effective Lagrangian approach, and the phenomenological
consequences have been extensively studied both at hadronic and
leptonic colliders \cite{Wudka}.

In this work we are interested in studying the on-shell $WW\gamma$
vertex in the framework of the minimal $331$ model, which is based
on the simplest nonabelian gauge-group extension of the SM, namely
$\mathrm{SU_c(3)}\times \mathrm{SU_L(3)}\times \mathrm{U_X(1)}$
\cite{Pleitez1}. In particular, we will concentrate on the
contributions coming from a pair of singly and doubly charged
gauge bosons predicted by this model. These particles are called
bileptons \footnote{Unless stated otherwise, throughout this work
we will use the terms bilepton or bilepton gauge boson to refer to
both the singly and doubly charged gauge bosons of the 331 model.}
because they have two units of lepton number. The 331 model has
attracted considerable attention recently \cite{Toscano} since it
requires that the number of fermion families be a multiple of the
quark color number in order to cancel anomalies, which offers a
possible solution to the flavor problem. Another important feature
of this model is that the $\mathrm{\mathrm{SU_L(2)}}$ group is
totally embedded in $\mathrm{SU_L(3)}$. As a consequence, after
the first stage of spontaneous symmetry breaking (SSB), when
$\mathrm{SU_L(3)}\times \mathrm{U_X(1)}$ is broken down to
$\mathrm{SU_L(2)}\times \mathrm{U_Y(1)}$, there emerge a pair of
massive bileptons in a doublet of the electroweak group, giving
rise to very interesting couplings with the SM gauge bosons. In
particular, these couplings do not involve any mixing angle, as
occurs in other SM extensions, and are similar both in strength
and Lorentz structure to those couplings existing between the SM
gauge bosons.

Besides their nontrivial transformation properties under the
electroweak gauge group, the bileptons get a mass splitting at the
Fermi scale due to the presence of some terms that violate the
custodial $\mathrm{SU_c(2)}$ symmetry. It is well known, from the
analysis of fermion or scalar doublets, that these peculiarities
might give rise to nondecoupling effects in low-energy processes.
For this reason, it is important to investigate the respective
contribution to $\Delta \kappa $ and $\Delta Q$ on the basis of
the decoupling theorem \cite{Appelquist}. It is a known fact that
a heavy particle might be detected through its virtual effects on
low-energy physics if it evades the decoupling theorem
\cite{nondecoupling}. This interesting phenomenon can occur only
in theories with SSB, where some particles can have a mass heavier
than the vacuum expectation value (VEV) scale due to a large
coupling constant. In such a situation, the suppression factor
arising from the propagator of the heavy particle is compensated
by a mass factor appearing in the numerator, which in turn is
determined by a large coupling constant. In contrast, a particle
decouples in the heavy mass limit if its mass is induced by a
gauge singlet bare parameter (usually an unfixed VEV) since
dynamics compensatory effects are not present in this case. In
this paper we will show that bileptons obey the decoupling theorem
and, as a consequence, their contribution to both $\Delta \kappa $
and $\Delta Q$ vanishes in the heavy mass limit. This behavior is
a result of the fact that a large mass implies a large parameter
not fixed by experiment, namely a VEV larger than the electroweak
scale. It is interesting to note that this case is similar to that
studied in Ref. \cite{Li}, in which an extra scalar doublet that
does not develop a VEV was considered. On the other hand, the
decoupling nature of $\Delta Q$ is not surprising, even if it
receives contributions from a particle that violates the
decoupling theorem. It turns out that this quantity is insensitive
to a large physical scale \cite{Inami}. This result is a
consequence of the fact that $\Delta Q$ is parametrized by a
dimension-6 Lorentz structure which is naturally suppressed by
inverse powers of the mass of the heavy particle circulating in
the loop, as was explicitly verified for the contribution of an
extra fermion generation and technihadrons \cite{Inami}. We will
return to this point later in the context of the bilepton
contribution.

In contrast to other extensions of the SM, in the $331$ model the
mass of the extra gauge bosons is bounded from above as a
consequence of matching the gauge coupling constants at the Fermi
scale \cite{DNg1}. Therefore this model would be either confirmed
or ruled out at the future high-energy colliders. Current bounds
establish that bilepton masses may take values ranging from a few
hundred of GeVs to about 1 TeV. This is an important reason to
investigate the effect of these particles on the $WW\gamma$
vertex. We will show below that the respective contributions to
$\Delta \kappa $ and $\Delta Q$ are comparable to those induced by
other weakly coupled renormalizable theories.

Another point worth to mention concerns the approach we took to
perform our calculation. In the first place, we found convenient
to use a nonlinear $R_\xi$ gauge rather than the unitary gauge.
For this aim we introduced a gauge-fixing term covariant under the
$\mathrm{U_e(1)}$ gauge group, from which the necessary Feynman
rules were derived. This gauge-fixing procedure allowed us to
remove the mixed $YG_Y\gamma$ vertices. As for the evaluation of
the tensorial integrals, it has been customary to use the Feynman
parameters technique for evaluating the static properties of
elementary particles. It turns out that, in this case, the
Passarino-Veltman reduction method \cite{Veltman1} breaks down
since the Gram determinant of the kinematic matrix vanishes
\cite{Stuart}. However, we will show below that, even in this
case, the last method can be used after introducing some slight
modifications.

The rest of the paper is organized as follows. In Sec. \ref{model}
we give a description of the minimal $331$ model. Particular
emphasis is given to the Yang-Mills sector. In Sec.
\ref{calculation} we present the calculation of the static
properties of the $W$ boson. Sec. \ref{discussion} is devoted to
discuss our results, and the conclusions are presented in Sec.
\ref{remarks}. Finally, explicit expressions for both the
trilinear and quartic vertices involving bilepton gauge bosons are
presented in the Appendices, together with the respective Feynman
rules.

\section{Review of the minimal $331$ model}
\label{model}

To begin with, we present a short description of the fermionic
sector of the minimal $331$ model. We will turn next to discuss in
detail the gauge sector. In particular, we will focus on the mass
spectrum and the coupling structure of the Yang-Mills sector.
Hereafter, we will follow closely the notation and conventions of
Ref. \cite{DNg2}. The simplest anomaly-free fermionic content of
the 331 model accommodates the leptons as antitriplets of
$\mathrm{SU_L(3)}$:

\begin{equation}
\ell^i_L= \left( \begin{array}{c} e^i_L \\
\nu^i_L \\
e^{c\,i}
\end{array} \right): (1, 3^*,0),
\end{equation}

\noindent where $i=1,2,3$ is the generation index. The quark
sector includes three new exotic quarks. Two quark generations are
given the same representation, and the third one (no matter what)
is treated differently. It has been customary to represent the
first two quark families as triplets and the third one as
antitriplet \cite{Pleitez1}:

\begin{equation}
q_L^i=
\left(\begin{array}{c}
u^{i}_L\\
d^{i}_L\\
D^{i}_L
\end{array}\right)
:~(3,3,-1/3),
\end{equation}
\begin{equation}
u^i_R:~(3,1,-2/3);\quad
d^i_R:~(3,1,+1/3);\quad
D^i_R:~(3,1,+4/3);\qquad (i=1,2),
\end{equation}

\begin{equation}
q^3_L=
\left(\begin{array}{c}
u^3_L\\
d^3_L\\
T^3_L
\end{array}\right):~(3,3^*,2/3),
\end{equation}
\begin{equation}
u^3_R:~(3,1,+1/3);\quad
d^3_R:~(3,1,-2/3);\quad
T^3_R:~(3,1,-5/3).
\end{equation}

In order to accomplish the gauge hierarchy and the fermion masses,
a Higgs sector composed of several $\mathrm{SU_L(3)}$ multiplets
is required: to break $\mathrm{SU_L(3)}\times \mathrm{U_X(1)}$
down to $\mathrm{SU_L(2)}\times \mathrm{U_Y(1)}$ only one
$\mathrm{SU_L(3)}$ scalar triplet is necessary; the next stage of
SSB, $\mathrm{SU_L(2)}\times \mathrm{U_Y(1)}\to \mathrm{U_e(1)}$,
requires two scalar $\mathrm{SU_L(3)}$ triplets and one sextet.
The minimal Higgs sector has the following quantum numbers

\begin{equation}
\phi_Y= \left( \begin{array}{c}
\Phi_Y \\
\phi^0
\end{array} \right): \quad (1,3,1); \quad
\phi_1= \left( \begin{array}{c}
\Phi_1 \\
\delta^-
\end{array} \right): \quad (1,3,0); \quad
\phi_2= \left( \begin{array}{c}
\widetilde{\Phi}_2 \\
\rho^{--}
\end{array} \right): \ \ \ (1,3,-1),
\end{equation}

\begin{equation}
H= \left( \begin{array}{ccc}
T & \widetilde{\Phi}_3/\sqrt{2}\\
\widetilde{\Phi}^T_3/\sqrt{2} & \eta^{--}
\end{array} \right): \ \ \ (1,6,0),
\end{equation}

\noindent where $T$ is a $2\times 2$ matrix given by

\begin{equation}
T= \left( \begin{array}{ccc}
T^{++} & T^+/\sqrt{2}\\
T^+/\sqrt{2} & T^0
\end{array} \right),
\end{equation}

\noindent Both $\Phi_Y$ and $\Phi_i$ ($i=1,\,2,\,3$ and
$\widetilde{\Phi}_i=i\,\tau^2\,\Phi^*_i$) are two-component  complex
quantities. We will see below that after the first stage of SSB
all these quantities constitute a specific representation of the
electroweak group. When $\phi_Y$ develops a VEV,
$\mathrm{SU_L(3)}\times \mathrm{U_X(1)}$ breaks down to
$\mathrm{SU_L(2)}\times \mathrm{U_Y(1)}$ and the exotic quarks and
the new gauge bosons acquire masses. The remaining multiplets
endow the SM particles with mass.

The covariant derivative in the fundamental representation of
$\mathrm{SU_L(3)}\times \mathrm{U_X(1)}$ can be written as

\begin{equation}
{D}_\mu=\partial_\mu-i\,g\,\frac{\lambda^a}{2}
\,W^a_\mu-i\,g_XX\,\frac{\lambda^9}{2}\,X_\mu, \qquad (a=1 \ldots
8),
\end{equation}

\noindent with $\lambda^a$ the Gell-man matrices and
$\lambda^9=\sqrt{2/3}\;{\mathrm diag(1,1,1)}$. The generators are
normalized according to $\mathrm{Tr}\,\lambda^a
\lambda^b=2\,\delta^{ab}$, which means that
$\mathrm{Tr}\,\lambda^9 \lambda^9=2$. The first stage of SSB is
accomplished by the VEV of $\phi_Y$,
$\phi^\dag_{Y0}=(0,0,u/\sqrt{2})$, according to the following
scheme: six generators are broken, namely $\lambda^b \phi_{Y0} \ne
0$ ($b=4 \ldots 9$), whereas the remaining ones leave invariant
the vacuum, namely $\lambda^a \phi_{Y0}=0$ ($a=1,\,2,\,3$). Notice
that $\sqrt{3}\,(\lambda^8+\sqrt{2}X\lambda^9)\phi_{Y0}=0$, so the
hypercharge can be identified with a linear combination of broken
generators as follows
$Y=\sqrt{3}\,(\lambda^8+\sqrt{2}X\lambda^9)$. At this stage of
SSB, there appears one pair of singly and doubly charged
bileptons, which are defined by

\begin{equation}
Y^{++}_\mu =\frac{1}{\sqrt{2}}\left(W^4_\mu-iW^5_\mu\right),
\qquad Y^+_\mu =\frac{1}{\sqrt{2}}\left(W^6_\mu-iW^7_\mu\right),
\end{equation}

\noindent and get a mass given by

\begin{equation}
\label{Ymass1} M_Y=\MY=\MX=\frac{g\,u}{2}.
\end{equation}

\noindent According to the quantum number assignment, the bilepton
gauge bosons fill out one doublet of $\mathrm{SU_L(2)}\times
\mathrm{U_Y(1)}$ with hypercharge $3$:

\begin{equation}
Y_\mu= \left(\begin{array}{c}
 Y^{++}_\mu \\ Y^+_\mu \end{array} \right).
\end{equation}

The gauge fields $W^8_\mu$ and $X_\mu$ mix to produce a massive
field, ${Z^\prime}_\mu$,  and a massless gauge boson, $B_\mu$. The
latter is associated with the $\mathrm{U_Y(1)}$ group. These
fields are given by

\begin{align}
{Z^\prime}_\mu&=c_\theta\, W^8_\mu-s_\theta\, X_\mu, \\
B_\mu&=s_\theta W^8_\mu+c_\theta X_\mu, \\
M^2_{{Z^\prime}}&=\frac{1}{6}\left(2\,g^2+g^2_X\right)u^2, \\
M_B&=0,
\end{align}

\noindent where $s_\theta=\sin\theta, c_\theta=\cos\theta$ and
$\tan\theta=g_X/(\sqrt{2}\,g)$. The coupling constant associated
with the hypercharge group is given by $g'=g\,s_\theta/\sqrt{3}$.
The remaining fields associated with the unbroken generators of
$\mathrm{SU_L(3)}$ are the gauge bosons of the
$\mathrm{\mathrm{SU_L(2)}}$ group, which will be denoted as
$W^i_\mu$ ($i=1,2,3$).

In the Higgs sector, $\Phi_Y$ and $\Phi_i$ ($i=1,2$) are
$\mathrm{\mathrm{SU_L(2)}}$ doublets with hypercharge $3$ and $1$,
respectively. It can be shown that the two components of $\Phi_Y$
represent the pseudo-Goldstone bosons associated with the bilepton
fields, while the real and imaginary part of $\phi^0$ correspond
to a physical Higgs boson and the pseudo-Goldstone boson
associated with the ${Z^\prime}$ field, respectively. The third
components of $\phi_1$ and $\phi_2$, $\delta^-$ and $\rho^{--}$,
are singlets of $\mathrm{\mathrm{SU_L(2)}}$ with hypercharge $-2$
and $-4$, respectively. The sextet $H$ is composed of the
following structures: a doublet $\Phi_3$ with $Y=+1$, a triplet
$T$ with $Y=+2$, and a singlet $\eta^{--}$ with $Y=+4$. As for the
fermionic sector, in addition to the SM content of leptons and
quarks, there appear three exotic quarks as singlets of
$\mathrm{\mathrm{SU_L(2)}}$. Among these exotic quarks, two of
them have electric charge $-4/3$, while the third one has electric
charge $5/3$.

In summary, after the first stage of SSB we end up with the \sm~
gauge group and the SM content of leptons and quarks, plus three
doublets and one triplet of scalar fields, one bilepton doublet,
one neutral gauge boson (${Z^\prime}$), and several singlets of
scalar and quark fields. The presence of a bilepton doublet is a
remarkable feature of this class of models and may give rise to
some interesting phenomenological consequences. The main aim of
this work is to explore the effects of these exotic particles on
the static electromagnetic properties of the $W$ gauge boson. The
respective contributions to the $WW\gamma$ vertex are dictated
entirely by the Yang-Mills sector of the model. It is interesting
to note that the exotic quarks do not contribute to the $WW\gamma$
vertex since they are $\mathrm{\mathrm{SU_L(2)}}$ singlets and
thus do not interact with the $W$ boson. Moreover, we will not
consider the contributions from charged scalar Higgs boson as such
kind of contributions have been studied widely within the two
Higgs doublet model \cite{Couture2}. Therefore, our main concern
lies on the structure of the Yang-Mills sector associated with the
$\mathrm{SU_L(3)}\times \mathrm{U_X(1)}$ group. The analysis of
the Higgs kinetic-energy terms is also required since this sector
is responsible for the splitting between the bilepton masses. As
it will be discussed below, such a splitting is a consequence of
the violation of the custodial $\mathrm{SU_c(2)}$ symmetry. In
addition, this sector requires some manipulation as we found
convenient to use a renormalizable $R_\xi$ gauge for our
calculation.

The full Yang-Mills Lagrangian is composed of the following three
$\mathrm{SU_L(2)}\times \mathrm{U_Y(1)}$ invariant pieces

\begin{align}
{\cal L}_{\mathrm{YM}}&=-\frac{1}{4}W^a_{\mu \nu}W^{\mu
\nu}_a-\frac{1}{4}X_{\mu \nu}X^{\mu \nu}, \nonumber \\
&={\cal L}_{\mathrm{SM}}+{\cal L}_{{\mathrm{SMNP}}}+{\cal
L}_{{\mathrm{NP}}},
\end{align}

\noindent where ${\cal L}_{\mathrm{SM}}$ is the SM Yang-Mills
Lagrangian given by

\begin{equation}
{\cal L}_{\mathrm{SM}}=-\frac{1}{4}W^i_{\mu \nu}W^{\mu
\nu}_i-\frac{1}{4}B_{\mu \nu}B^{\mu \nu}.
\end{equation}

\noindent ${\cal L}_{{\mathrm{SMNP}}}$, which comprises the
interactions between the SM gauge bosons and  the new ones, can be
written as

\begin{align}
\label{LSMNP}
 {\cal L}_{{\mathrm{SMNP}}}=&-\frac{1}{2}\left(D_\mu
Y_\nu-D_\nu Y_\mu\right)^\dag \left(D^\mu Y^\nu-D^\nu Y^\mu\right) -Y^{\dag
\mu}\left(i\,g\,{\bf W}_{\mu \nu}+i\,g'\,{\bf B}_{\mu \nu}\right)Y^\nu
\nonumber \\
&-\frac{i\,\sqrt{3}\,g\,c_\theta}{2}{Z^\prime}_\mu \left[Y^\dag_\nu
\left(D^\mu Y^\nu-D^\nu Y^\mu\right)-\left(D^\mu Y^\nu-D^\nu Y^\mu\right)^\dag Y_\nu\right]
\end{align}

\noindent where we have introduced the definitions ${\bf W}_{\mu
\nu}=\tau^i \, W^i_{\mu \nu}/2$ and ${\bf B}_{\mu \nu}=Y\,B_{\mu
\nu}$/2. In addition $D_\mu=\partial_\mu -ig{\bf W}_\mu -ig'{\bf
B}_\mu$ is the covariant derivative associated with the
electroweak group. Such a Lagrangian induces new couplings, which
possess a rich structure, between the SM gauge bosons and the
bileptons. It is interesting to note that the ${Z^\prime}WW$
vertex is not induced. In particular, the trilinear vertices
$WYY$, $YY\gamma$, and the quartic one $WWYY$, induce one-loop
anomalous contributions to the electromagnetic static properties
of the $W$ boson. The trilinear couplings were previously studied
in \cite{Long}. We take one step forward and present the complete
expressions for both the trilinear and the quartic vertices in
Appendix \ref{AppendixI}. Finally, the term ${\cal
L}_{{\mathrm{NP}}}$ induces interactions between the ${Z^\prime}$
boson and the bileptons:

\begin{align}
{\cal L}_{{\mathrm{NP}}}=&-\frac{1}{4}{Z^\prime}_{\mu
\nu}{Z^\prime}^{\mu
\nu}-\frac{\sqrt{3}\,g\,c_\theta}{2}{Z^\prime}_{\mu \nu}Y^{\dag
\mu}Y^\nu-\frac{3\,g^2\,c^2_\theta}{4}{Z^\prime}_\mu
Y^\dag_\nu\left({Z^\prime}^\mu Y^\nu-{Z^\prime}^\nu Y^\mu\right)\nonumber \\
&+\frac{g^2}{2}\left(Y^\dag_\mu \,\frac{\tau^i}{2}Y_\nu\right)\left(Y^{\dag
\mu}\,\frac{\tau^i}{2}Y^\nu-Y^{\dag
\nu}\frac{\tau^i}{2}Y^\mu\right)+\frac{3\,g^2}{4}\left(Y^\dag_\mu
Y_\nu\right)\left(Y^{\dag \mu}Y^\nu-Y^{\dag \nu}Y^\mu\right).
\end{align}

At the Fermi scale the bileptons and the ${Z^\prime}$ boson
receive additional mass contributions from the VEV of the
$\mathrm{\mathrm{SU_L(2)}}$ doublet $<\Phi^0_i>_0=v_i/\sqrt{2}$
($i=1,2$). By simplicity we are assuming that $<H>_0=0$. The extra
mass terms for the bileptons arise from the Higgs kinetic-energy
sector and are given by

\begin{equation}
V=\frac{g^2}{2}\left[\left(Y^\dag_\mu \Phi_1\right)\left(\Phi^\dag_1 Y^\mu\right)+
\left(Y^\dag_\mu\widetilde{\Phi}_2\right)\left(\widetilde{\Phi}^\dag_2 Y^\dag\right)
 \right].
\end{equation}

\noindent Notice that these terms violate the custodial $\mathrm
SU_c(2)$ symmetry. Therefore, the bilepton masses are now given by

\begin{equation}
\MY=\frac{g^2}{4}\left(u^2+v^2_2\right), \quad
\MX=\frac{g^2}{4}\left(u^2+v^2_1\right). \label{Ymass1a}
\end{equation}

\noindent As for the $W$ boson, it gains a mass given by

\begin{equation}
m^2_W=\frac{g^2}{4}\left(v^2_1+v^2_2\right).
\end{equation}

\noindent From these expressions, the following bound on the
splitting between the squared bilepton masses can be derived
\cite{DNg3}

\begin{equation}
\left|\SMY-\SMX\right|\leq m^2_W.
\label{Ymass2}
\end{equation}

Some remarks concerning the decoupling theorem and the custodial
symmetry are in order. First of all, note that the bilepton masses
depend essentially on the coupling constant $g$, the Fermi scale
$v^2=v^2_1+v^2_2$, and $u$. Since $g$ and $v$ are fixed by
experiment, the only way in which the bileptons can become very heavy
is through a large $u$. We will discuss below that this fact is
crucial in order for the bileptons to respect the decoupling
theorem. On the other hand, the mass splitting arises from the
term that violates the custodial $\mathrm{SU_c(2)}$ symmetry. As
an immediate consequence, there are bilepton contributions to the
$S\,T\,U$ oblique parameters arising from the mass splitting
\cite{DNg3}. As it will be seen below, both $\Delta \kappa$ and
$\Delta Q$ also depend on this quantity, though the dependence is
somewhat different. In the large mass limit ($M^2_Y\gg m^2_W$) the
custodial symmetry is restored, i.e. $T \to 0$.

We can now specify the theory by defining a supplementary
condition. Since we are interested in studying the loop
contributions to the on-shell $WW\gamma$ vertex arising from
bileptons, it is only necessary to define a gauge-fixing procedure
for these fields. Although a calculation within the unitary gauge
requires only those vertices which arise from ${\cal
L}_{{\mathrm{SMNP}}}$, for computational matters we found
convenient to work in the framework of a renormalizable $R_\xi$
gauge, which requires the introduction of scalar fields
(pseudo-Goldstone bosons and Faddeev-Popov ghosts). We will define
a gauge which is covariant under the electromagnetic
$\mathrm{U_e(1)}$ group by means of gauge-fixing functions which
transform covariantly under this group \cite{Fujikawa}. We
summarize the respective gauge-fixing procedure, together with the
Feynman rules necessary for our calculation in Appendix
\ref{AppendixII}.

\section{Static properties of the $W$ boson}
\label{calculation}

\begin{figure}
\centerline{\epsfig{file=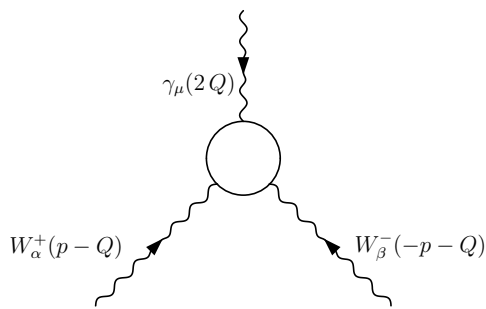,width=2.5in,clip=}} \caption{The
trilinear $WW\gamma$ vertex. The loop denotes any anomalous
contribution.} \label{Fig-WWg}
\end{figure}

When the three bosons are on the mass-shell, the most general $CP$-conserving
$WW\gamma$ vertex can be written as \cite{Bardeen}

\begin{align}
\Gamma^{\mu \alpha \beta}&=i\,e\bigg\{A\left[2\,p^\mu g^{\alpha
\beta}+4\,\left(Q^\beta\,g^{\mu \alpha}-Q^\alpha\,g^{\mu
\beta}\right)\right]\nonumber \\
&+2\,\Delta \kappa \left(Q^\beta\,g^{\mu \alpha}-Q^\alpha\,g^{\mu
\beta}\right)+\frac{4 \,\Delta Q }{m_W^2}p_\mu
\,Q^\alpha\,Q^\beta\bigg\}, \label{WWg}
\end{align}

\noindent where we are using the set of variables depicted in Fig.
\ref{Fig-WWg}. In the SM, both $\Delta \kappa$ and $\Delta Q$
vanish at tree level, whereas the one-loop corrections are of the
order of $\alpha/\pi$ \cite{Bardeen}. These parameters are defined
as

\begin{equation}
\Delta \kappa=\kappa_\gamma+\lambda_\gamma-1,
\end{equation}
\begin{equation}
\Delta Q=-2\,\lambda_\gamma.
\end{equation}

\noindent where $\kappa_\gamma$ and $\lambda_\gamma$ are related
in turn to the magnetic dipole moment, $\mu_W$, and the
electric quadrupole moment, $Q_W$, as follows

\begin{equation}
\mu_W=\frac{e}{2\,m_W}\,\left(1+\kappa_\gamma+\lambda_\gamma\right),
\end{equation}

\begin{equation}
Q_W=-\frac{e}{m_W^2}\,\left(\kappa_\gamma-\lambda_\gamma\right).
\end{equation}

\begin{figure}
\centerline{\epsfig{file=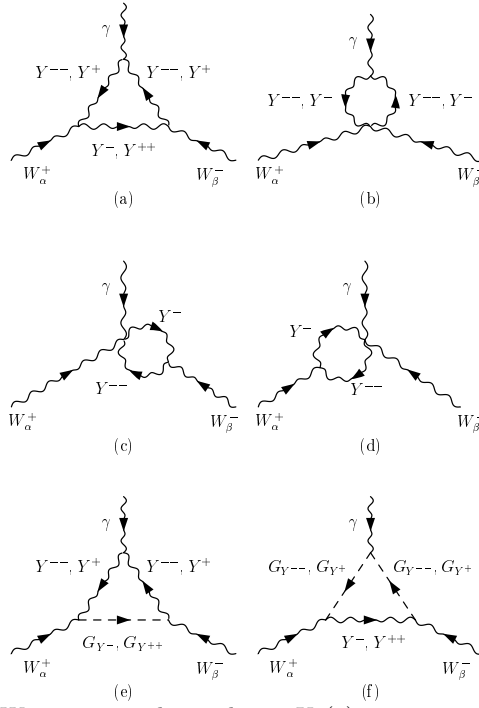,width=2.5in,clip=}}
\caption{Feynman diagrams for the $WW\gamma$ vertex in the
nonlinear $\mathrm{U_e(1)}$-covariant gauge. There are also two
sets of diagrams obtained from diagrams (a)-(d) after replacing
each bilepton gauge boson with their respective pseudo-Goldstone
boson and diagrams (a), (c) and (d) with their ghost field.}
\label{Fig-WWgFD}
\end{figure}

In this Section we will present the complete calculation of the
bilepton contribution to both $\Delta Q$ and $\Delta \kappa$ in
the minimal $331$ model. Before presenting our results, it is
worth commenting about the scheme that was employed to calculate
the contribution from the diagrams of Fig. \ref{Fig-WWgFD}.

In the nonlinear $\mathrm{U_e(1)}$-covariant gauge, the static
properties of the $W$ boson receive contributions from singly and
doubly charged bileptons through the diagrams depicted in Fig.
\ref{Fig-WWgFD}. In the Feynman-t' Hooft gauge, that can be safely
used since the static properties are gauge independent, there are
also contributions from diagrams with unphysical fields. Apart
from diagrams \ref{Fig-WWgFD}(e) and \ref{Fig-WWgFD}(f), the
singly and doubly charged pseudo-Goldstone bosons also contribute
through two triangle diagrams similar to that shown in Fig.
\ref{Fig-WWgFD}(a). It is easy to see that there are no
contributions from any two-point diagram involving only
pseudo-Goldstone bosons. The same is true for the singly and
doubly charged Ghost field contributions, which arise only from
triangle diagrams similar to that of Fig. \ref{Fig-WWgFD}(a).
Given the Feynman rules shown in Appendix \ref{AppendixII}, it is
straightforward to obtain the amplitude corresponding to each one
of the diagrams contributing to the static properties of the $W$
boson. We have used a slightly modified version of the reduction
scheme of Passarino and Veltman to express our result in terms of
scalar functions \cite{Veltman1}. To illustrate our calculation
scheme, let us consider the triangle diagram shown in Fig.
\ref{Fig-WWgFD} (a): the one which involves the
$Y^{--}Y^{--}\gamma$ coupling but now with the bilepton gauge
bosons replaced with their respective pseudo-Goldstone boson. From
now on this diagram will be referred to as $(a^\prime)$. The
respective amplitude is given by

\begin{equation}
{\mathcal{M}}^{(a^\prime)}_{\alpha \beta \mu}=8\,g^2\,e \int
\frac{d^Dk}{(4\pi)^D}\frac{(k+Q)_\alpha (k-Q)_\beta
k_\mu}{\Delta},
\end{equation}

\noindent where

\begin{equation}
\Delta=\left((k+p)^2-\SMX\right)\left((k+Q)^2-\SMY\right)\left((k-Q)^2-\SMY\right),
\end{equation}

\noindent and $D$ is the space-time dimension. We have dropped any
term that does not contribute to the static properties of the $W$
boson. In addition, we have used the mass-shell and transversality
conditions for the gauge bosons, which means that we can make the
following replacements everywhere: $(Q-p)^2=(Q+p)^2=m_W^2$,
$p^\alpha \to Q^\alpha$ and $p^\beta \to -Q^\beta$. As it will be
evident below, the condition $Q^2=0$ was only used at the final
stage of the calculation.

The amplitude ${\mathcal{M}}^{(a^\prime)}_{\alpha \beta \mu}$ can
be put in the form of Eq. (\ref{WWg}) by means of the Feynman
parameters technique. One more alternative is to use the
Passarino-Veltman method to reduce the tensor integrals down to
scalar functions. However, the last scheme involves the inversion
of the kinematic matrix

\begin{equation}
{\mathbf{\cal D}}=\left|
\begin{array}{cc}
p_1^2& p_1\cdot p_2\\
p_1\cdot p_2&p_2^2
\end{array}
\right|,
\end{equation}

\noindent where $p_1=Q-p$ and $p_2=-2\,Q$. The respective Gram
determinant is given by $\|{\mathbf{\cal
D}}\|=4\,Q^2\left(m^2_W-Q^2\right)$, which clearly vanishes for
$Q^2=0$. It is thus evident that the Passarino-Veltman method
breaks down if one attempts to use the condition $Q^2=0$ during
the course of the reduction stage. One way to overcome this
difficulty is via the approach followed in Ref. \cite{Stuart},
which can be summarized in two steps

\begin{itemize}
\item Assume that $Q^2\neq 0$ and apply the Passarino-Veltman reduction scheme as
usual.
\item Once the reduction is done, take the limit $Q^2\to 0$.
\end{itemize}

After the reduction stage, the contribution from the diagram
(a$^\prime$) to $\Delta Q$ can be expressed as

\begin{align}
\frac{16\,\pi^2\,\Delta Q^{(a^\prime)}}{g^2}&=\lim_{Q^2\to
0}\,\bigg[ \frac{1}{m_W^2\left(m_W^2-Q^2\right)^3}
\bigg(\frac{1}{Q^2}h_0(Q^2)+h_1(Q^2) \bigg)\bigg]. \label{deltaQ1}
\end{align}

\noindent Both the $h_0$ and $h_1$ functions are analytical at
$Q^2=0$. In fact $h_0(0)=0$, which is a necessary condition in
order for the limit of Eq. (\ref{deltaQ1}) to exist. This function
is given in terms of scalar functions as follows

\begin{align}
h_0(Q^2)&=\beta_0+\beta_1\, B_0(0,\SMY,\SMY)+\beta_2
\,B_0(0,\SMX,\SMX)+ \beta_3\,B_0(m_W^2,\SMX,\SMY)
\nonumber\\&+\beta_4\,B_0(Q^2,\SMY,\SMY) +\beta_5\,
C_0\left(m_W^2,m_W^2,4Q^2,\SMY,\SMX,\SMY\right),
\end{align}

\noindent  where  the $\beta_i$ functions depend on $Q^2$,
$m_W^2$, $\SMX$, and $\SMY$. A similar expression holds for the
$h_1$ function. We are using the notation of Ref. \cite{Mertig}
for the scalar functions. The application of l'H\^opital rule to
Eq. (\ref{deltaQ1}) yields

\begin{align} \Delta
Q^{(a^\prime)}&=\frac{\alpha}{4 \pi s_W^2 m_W^8}
\bigg(\frac{\partial h_0(Q^2)}{\partial Q^2}\bigg|_{Q^2=0}+h_1(0)
\bigg). \label{deltaQ2}
\end{align}

As was noted in  Ref. \cite{Stuart}, any $n$-point scalar function
and its respective derivatives can be expressed in terms of a set
of $(n-1)$-point scalar functions when the kinematic Gram
determinant vanishes. It follows that, in the limit of $Q^2 \to
0$, one can express the three-point scalar function
$C_0\left(m_W^2,m_W^2,4\,Q^2,\SMY,\SMX,\SMY\right)$ and its
derivative with respect $Q^2$ in terms of the two-point scalar
functions $B_0\left(m^2_W,\SMX,\SMY \right)$,
$B_0\left(0,\SMX,\SMX \right)$, and $B_0\left(0,\SMY,\SMY
\right)$. It is then straightforward, though somewhat lengthy, to
obtain the limit of Eq. (\ref{deltaQ1}). We thus have

\begin{align}
\Delta Q^{(a^\prime)}&=-\frac{g^2}{24\, \pi^2\,{\zeta }^2
}\bigg\{\frac{1}{3}\left[ \left(2 - 3\,\eta \, \left( 3 + 2\,\eta
\right)  + 3\,\xi  + 12\,\eta \,\xi  - 6\,{\xi }^2\right)\,{\zeta
}^2 - 12\,\eta \,\left( 1 - \eta  - \xi \right)
 \right]\nonumber \\
&+2\left[\left(\eta \,\left( 1 + \eta  \right)  - \left( 1 +
2\,\eta \right) \,\xi  + {\xi }^2\right)\,{\zeta }^2 + \eta
\,\left( 1 - \eta
+ \xi  \right) \right] F_1(\eta,\xi)\nonumber \\
&-2\left( 1 + \eta  - \xi  \right) \,\xi \, \left(\eta-{\zeta }^2
\right) F_2(\eta,\xi) \bigg\},
\end{align}

\noindent where we are using the following variables
$\eta=\left(\MX/m_W\right)^2$, $\xi=\left(\MY/m_W\right)^2$,
$\zeta^2=4\,\eta-(\xi-\eta-1)^2$, and $\omega^2=4\, \eta\, \xi$.
Furthermore

\begin{align}
F_1(\eta,\xi)&=B_0(m_W^2,\SMX,
\SMY)-B_0(0,\SMX,\SMX)\nonumber \\
&=\frac{1}{2}\left[4 - 2\,\zeta\, \arcsin \left(\frac{\zeta
}{\omega }\right) + \left( \xi - \eta  -1\right) \, \log
\left(\frac{\eta }{\xi }\right)\right],
\label{f1}
\end{align}
\begin{equation}
F_2(\eta,\xi)=B_0(0,\SMY,\SMY)-B_0(0,\SMX,\SMX)
=\log\left(\frac{\eta }{\xi }\right). \label{f2}
\end{equation}

\noindent We introduced explicit solutions for the scalar
two-point functions.

In a similar way, we can obtain the respective contribution to
$\Delta \kappa$, which is given by

\begin{align}
\Delta \kappa^{(a^\prime)}&=-\frac{g^2}{16\, \pi^2 }\bigg\{\eta
\left(1-4\,\xi+2\,\eta \right)- \xi \, \left( 1 - 2\,\xi \right) -
\frac{1}{3}  \nonumber \\
&-2\,\left( {\left( \eta  - \xi  \right) }^2 - \xi  \right)
F_1(\eta,\xi) -2\,\xi \,\left( 1 + \eta  - \xi
\right)F_2(\eta,\xi) \bigg\}.
\end{align}

The scheme above outlined can be employed to calculate the
contributions from the whole set of diagrams shown in Fig.
\ref{Fig-WWgFD}. Apart from the computational facilities offered
by this scheme \cite{Mertig}, its advantages are twofold: the
cancellation of ultraviolet divergences is evident [see Eqs.
(\ref{f1}) and (\ref{f2})] since they are isolated as poles of
$4-D$ in the two-point scalar functions; furthermore, the
two-point scalar functions in turn can be expressed in terms of
elementary functions \cite{Veltman2} or numerically evaluated
readily \cite{FF}.

We turn now to discuss some other interesting facts of our
calculation. In the nonlinear $\mathrm{U_e(1)}$-covariant gauge,
the only diagrams with ultraviolet divergences are those shown in
Fig. \ref{Fig-WWgFD}(a) - \ref{Fig-WWgFD}(d). The ultraviolet
divergences of diagram \ref{Fig-WWgFD}(a) cancel out those of
diagrams \ref{Fig-WWgFD}(b)- \ref{Fig-WWgFD}(d). It is interesting
to note that in the SM the ultraviolet divergences are cancelled
when the contribution from the two-point diagram with $W$ bosons
[the analogue of diagram \ref{Fig-WWgFD}(b)] is added to the
contribution from the triangle diagrams including a $Z$ boson or a
photon. In the case of the 331 model, there is no contribution
from the extra $Z^\prime$ boson and divergence cancellation occurs
between diagrams containing just bileptons.

As shown in Eq. (\ref{Ymass2}), in the minimal 331 model SSB
imposes an upper bound on the splitting of the bilepton masses,
which can be rewritten in terms of $\eta$ and $\xi$ as
$|\xi-\eta|\leq\, 1$. Therefore, it is convenient to express the
static properties of the $W$ boson in terms of one bilepton mass
and the mass splitting, let us say $\eta$ and
$\epsilon=\xi-\eta\leq 1$ (we have assumed that $\xi > \eta$). We
thus rewrite $\zeta$ and $\omega$ as $\zeta^2=4\, \eta -
{\left(\epsilon -1\right)^2}$ and $\omega^2=4\,\eta\, \left(\eta +
\epsilon \right)$. Once the contribution from the diagrams of Fig.
\ref{Fig-WWgFD} has been obtained, one can write

\begin{equation}
\Delta Q= \frac{\alpha}{4 \pi s_w^2}\left[f_0^Q+
\frac{1}{\zeta}\arcsin
\left(\frac{\zeta }{\omega }\right)f_1^Q +
\log\left(\frac{\eta}{\eta + \epsilon }\right)f_2^Q\right],
\label{deltaQgen}
\end{equation}

\noindent with

\begin{equation}
f_0^Q=\frac{2}{3}+\epsilon \,
\left(2\,\epsilon-7 \right)-2\,\eta,
\end{equation}

\begin{equation}
f_1^Q=2\,\left[\left(\epsilon-1  \right) \,\epsilon \,
\left(2+\left(\epsilon-4 \right) \,
\epsilon \right) -\left(1+\epsilon \,
\left(4\,\epsilon-11\right)\right) \,\eta  +
2\,{\eta }^2 \right],
\end{equation}

\begin{equation}
f_2^Q=\epsilon \,\left(2+\left(\epsilon-4\right) \,
\epsilon-2\,\eta \right) +3\,\eta.
\end{equation}

We also have

\begin{equation}
\Delta \kappa= \frac{\alpha}{4 \pi s_W^2}\bigg[f_0^\kappa+
\frac{1}{\zeta}\arcsin
\left(\frac{\zeta }{\omega }\right)f_1^\kappa+
\log\left(\frac{\eta}{\eta + \epsilon }\right)f_2^\kappa\bigg],
\label{deltakgen}
\end{equation}

\noindent with

\begin{equation}
f_0^\kappa=-\frac{3}{2}\left[3-\left(5-2\,\epsilon\right) \,
\epsilon-4\,\eta \right],
\end{equation}

\begin{equation}
f_1^\kappa=-2+\left(1-\epsilon\right) \,\epsilon \,
\left(7-3\,\left(\epsilon-3 \right)\,
\epsilon\right)+11\,\eta-
15\,\left(\epsilon-2\right)\,\epsilon\,\eta-
12\,{\eta }^2,
\end{equation}

\begin{align}
f_2^\kappa&=\frac{1}{2}\left[6-9\,\eta-\epsilon \,
\left(7-3\,\left(3-\epsilon\right) \,
\epsilon-9\,\eta\right)\right].
\label{f2keq}
\end{align}

A very interesting scenario is that in which the bilepton masses are
degenerate ($\epsilon=0$). From the above equations we get

\begin{equation}
\label{deltaQdeg}
\Delta Q\big|_{\epsilon=0}=
\frac{\alpha}{4 \pi s_W^2}\left(\frac{2}{3} - 2\,\eta -
\frac{2\,\eta\,\left( 1 - 2\,\eta \right) \, }{{\sqrt{
4\,\eta-1}}}\arcsin{\frac{{\sqrt{4\,\eta-1}}}{2\,\eta}}\, \right),
\end{equation}

\begin{equation}
\label{deltakdeg}
\Delta \kappa\big|_{\epsilon=0}=
\frac{\alpha}{4 \pi s_W^2}\left( -\frac{9}{2} + 6\,\eta +\left( 2
- 3\,\eta \right) \,{\sqrt{4\,\eta-1}}\, \arcsin{\frac{{\sqrt{
4\,\eta-1}}}{2\,\eta}}\,\right),
\end{equation}

We will discuss below the decoupling properties of these quantities as
$\eta \to \infty$.

\section{Results and discussion}
\label{discussion}

Before analyzing our results, it is interesting to comment on the
bounds on the doubly and singly charged gauge bosons masses. Both
scalar and vector bileptons have been extensively studied in the
literature \cite{bileptons}. An interesting peculiarity of the
minimal 331 model is that the masses of the bilepton gauge bosons
are bounded from above: $M_Y \leq 600$ GeV. This bound is derived
from the fact that embedding the electroweak group in the 331
gauge group requires that $\sin\theta_W < 1/4$
\cite{DNg3,Frampton}. However, it has been noted that the last
bound relaxes if the minimal Higgs sector is extended to comprise
one Higgs scalar octet or in other exotic scalar sectors
\cite{Frampton}. Up to now, the most stringent bound on the doubly
charged bilepton mass is that derived from muonium-antimuonium
conversion, $\mu^-e^+ \to \mu^+e^-$, which imposes the limit $\MY
\geq 800$ GeV \cite{Willman}. However, diverse authors have argued
that this bound can be evaded in a more general context since it
relies on some very restrictive assumptions, such as considering
that the matrix that couples bileptons to leptons is flavor
diagonal\cite{Pleitez2}. Another very stringent bound, namely $\MY
\geq 750$ GeV, arises from fermion pair production and lepton-flavor
violating processes \cite{Joshi}. Most recently, it has been
claimed that the data taken at the CERN LEP-II at $\sqrt
S=130-206$ GeV can be used to establish very restrictive bounds on
the doubly charged bilepton mass and the respective couplings
\cite{Gregores}. As for the singly charged gauge boson, the bound
$\MX>$ 440 GeV has been derived from limits on muon decay
parameters \cite{Joshi2}. However, we can directly obtain a bound
on the mass of this bilepton by considering the mass splitting
bound [Eq. (\ref{Ymass2})] and the current bounds on $\MY$. At
this point we would like to stress that all the previous bounds
are somewhat model dependent, thereby allowing the existence of a
lighter bilepton gauge boson. In the following analysis, we will
consider the more conservative range 300 GeV $\leq M_Y \leq $ 1
TeV. We will discuss below that the mass splitting bound has very
important consequences which are closely related to the decoupling
theorem.

\begin{figure}
\centerline{\epsfig{file=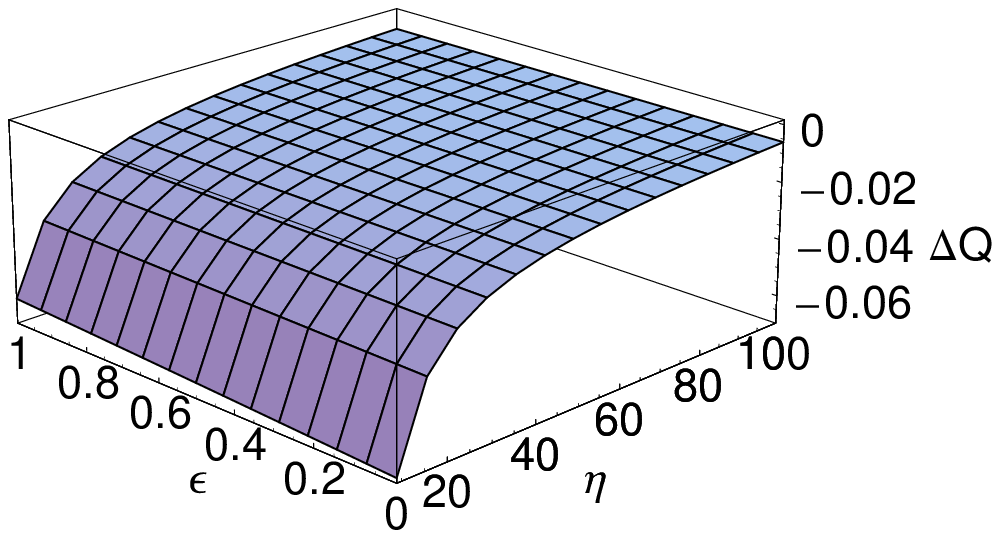,width=3in,clip=}}
\caption{Bilepton gauge boson contribution to the anomalous
$\Delta Q$ parameter, in units of $a=g^2/(96\,\pi^2)$.
$\eta=(\MX/m_W)^2$ and $\epsilon=(\SMY-\SMX)/m_W^2$.}
\label{Fig-deltaQa}
\end{figure}

\begin{figure}
\centerline{\epsfig{file=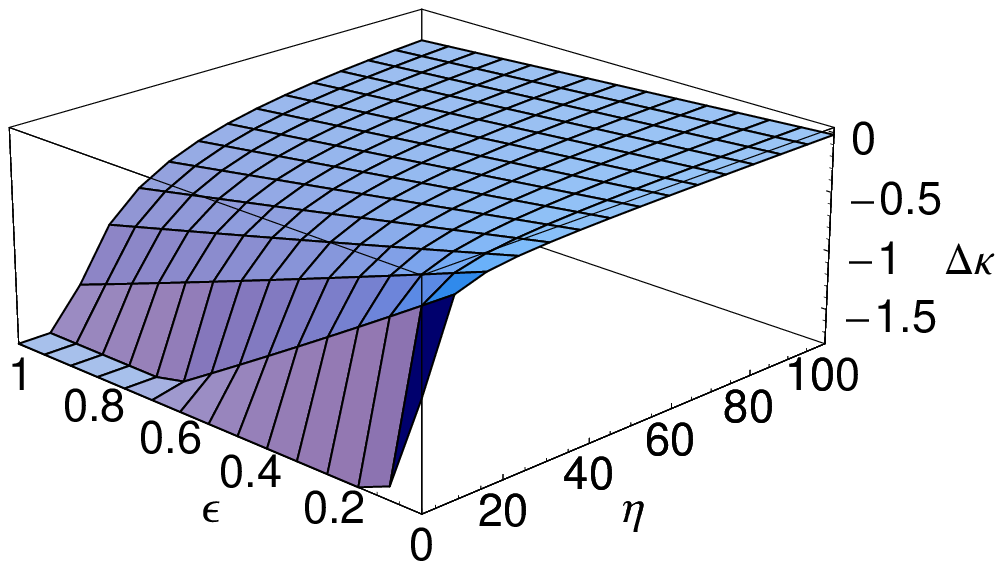,width=3in,clip=}}
\caption{Bilepton gauge boson contribution to the anomalous
$\Delta \kappa$, in units of $a=g^2/(96\,\pi^2)$.
$\eta=(\MX/m_W)^2$ and $\epsilon=(\SMY-\SMX)/m_W^2$.}
\label{Fig-deltaka}
\end{figure}

We are now ready to discuss our results. In the preceding Section we have presented
explicit expressions, ready for their numerical evaluation, of the bilepton
contribution to the static properties of the $W$ boson in the minimal 331 model. To
begin with, it is worth analyzing the behavior of the $\Delta Q$ and $\Delta \kappa$
parameters as functions of both the singly and doubly charged gauge boson masses. The
$\Delta Q$ parameter is shown in Fig. \ref{Fig-deltaQa}, whereas Fig.
\ref{Fig-deltaka} shows the $\Delta \kappa$ parameter. For the purpose of comparison
with results derived within other models, the results shown in Fig. \ref{Fig-deltaQa}
and Fig. \ref{Fig-deltaka} are given in units of $a=g^2/(96 \pi^2)$, which has been
widely used in the literature\cite{Couture1}. We are using the scaled variables
$\eta$, $\xi$, and $\epsilon$, defined in the last Section. It is not surprising that
the maximum contribution from the bilepton gauge bosons to $\Delta Q$ is of the order
of $O(a/100)$, while the maximum value of $\Delta \kappa$ is of the order of $O(a)$.
These values are of the same order of magnitude as those arising from other weakly
coupled renormalizable theories, such as the two-Higgs doublet model
\cite{Couture2}, supersymmetric theories \cite{SUSY}, etc. As far as the SM
contributions are concerned, $\Delta Q \simeq O(a/10)$ and $\Delta \kappa \simeq
O(10\,a)$, for a Higgs boson mass of the order of 100 GeV \cite{Couture1}. In the 331
model, the maximum value of $\Delta Q$ is reached when the bilepton gauge boson
masses are degenerate and acquire their lowest allowed values. In the case of $\Delta
\kappa$, its maximum value is obtained for the lightest allowed singly charged
bilepton and the maximum allowed splitting. Both $\Delta Q$ and $\Delta \kappa$
decrease rapidly as the bilepton masses increase simultaneously, as expected from the
decoupling theorem. We will argue below that the validity of the decoupling theorem
is, in this case, a little more involved than usual.

\begin{figure}
\centerline{\epsfig{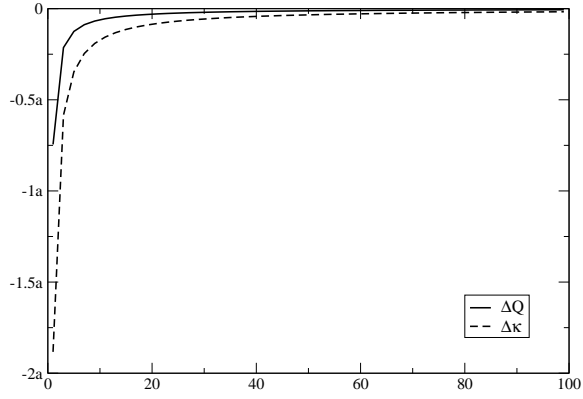}}
\caption{Static properties of the $W$ boson, in units of
$a=g^2/(96\,\pi^2)$, when the bilepton masses are exactly
degenerate, as a function $\eta=(M_Y/m_W)^2$.}
\label{Fig-decoupling}
\end{figure}

If the 331 model is realized in nature, there is no compelling
reason to expect that the bilepton masses are exactly degenerate.
However, in the case of a very heavy bilepton, with a mass of the
order of 1 TeV, the bilepton masses are indeed almost degenerate
(for instance, when $\MY=1$ TeV the maximum splitting allows for
$\MX\simeq 997$ GeV). Therefore, in the heavy mass limit the
bilepton masses become exactly degenerate and the custodial
$\mathrm{SU_c(2)}$ symmetry is also exact, which also means that
in this limit the bilepton contribution to the oblique parameter
$T$ vanishes \cite{DNg3}. In Fig. \ref{Fig-decoupling} we show the
static properties of the $W$ boson, as a function of $\eta$, when
the bilepton masses are degenerate. In this scenario, both $\Delta
Q$ and $\Delta \kappa$ decouple from low-energy physics when the
bilepton mass is very heavy, in accordance with the decoupling
theorem. It is interesting to analyze further this point. The
bilepton gauge bosons acquire masses from the VEV of $\Phi_Y$,
when the \suu~ gauge group is broken down to the \sm~ gauge group.
At this stage of SSB, the bilepton gauge boson masses are
degenerate [see Eq. (\ref{Ymass1})]. The subsequent breaking of
the \sm~ gauge group, through the VEVs of the $\Phi_i^0$ doublets,
induces the splitting $\epsilon$ between the bilepton masses [see
Eq. (\ref{Ymass1a})], thereby breaking the custodial
$\mathrm{SU_c(2)}$ symmetry. Since the SM gauge boson also get
their masses at this stage, $\epsilon$ cannot become arbitrarily
large and is bounded from above. In fact, a heavy bilepton mass
implies a large VEV of $\Phi_Y$, which is not fixed by experiment.
So, we cannot have a scenario where the mass of one bilepton
becomes large while the mass of the other one remains small. This
fact is crucial for the validity of the decoupling theorem. At
this point we would like to compare the bilepton case with that of
a SM-like fermion doublet, which is known to give rise to
nondecoupling effects when there is a large splitting between the
masses of the fermion doublet components \cite{Veltman3}. For this
purpose, let us consider the contribution from a hypothetical
SM-like fourth fermion family, with quarks $u^\prime$ and
$d^\prime$, to the oblique parameters. Of course, the same
analysis applies to a doublet composed of a heavy lepton and a
massive neutrino \cite{Li2}. In this case, the fermion masses
arise from independent Yukawa couplings and a heavy mass involves
a large Yukawa coupling. In principle, there is no theoretical
restriction for having a splitting $\Delta
m_f^2=m_{u^\prime}^2-m_{d^\prime}^2$ arbitrarily large, though
low-energy data do impose restrictions on it \cite{Veltman3}. To
clarify our point, let us consider the fermion contribution to the
$T$ oblique parameter:

\begin{equation}
T \sim F(m_{u^\prime}^2,m_{d^\prime}^2)=m_{u^\prime}^2+m_{d^\prime}^2-2\,
\frac{m_{d^\prime}^2\,m_{u^\prime}^2}{\Delta
m_f^2}\log\left(\frac{m_{u^\prime}^2}{m_{d^\prime}^2}\right),
\end{equation}

\noindent which clearly vanishes when $m_{u^\prime}=m_{d^\prime}$.
Let us now assume that we can make $m_{u^\prime}$ large while
$m_{d^\prime}$ is held fixed. In the limit $m_{d^\prime}\ll
m_{u^\prime}$ we get $T \sim m_{u^\prime}^2$. It is thus evident
that the decoupling theorem breaks down, which is not surprising
since the heavy mass limit implies a large Yukawa coupling.

Let us now consider the contribution from the bilepton gauge
bosons to the oblique $T$ parameter, which actually has the same
mass dependence as in the fermion case, i.e. $T\sim F(\SMX,\SMY)$
\cite{DNg3}. At first sight one might think that the bilepton
gauge bosons would also give rise to nondecoupling effects.
However, the splitting between the bilepton masses is now bounded
from above: $\Delta M_Y^2=\SMY-\SMX \leq m_W^2$, which in the
heavy mass limit becomes $\Delta M_Y^2\ll \SMY \sim \SMX$.
Therefore, writing $\SMY=\Delta M_Y^2-\SMX$ and expanding the $T$
parameter in powers of $\Delta m^2/\SMX$, we have in the limit of
large bilepton masses

\begin{equation}
T \sim \frac{{(\Delta M_Y}^2)^2}{\SMX} \sim \frac{{(\Delta
M_Y}^2)^2}{\SMY}.
\end{equation}

\noindent It is thus clear that in this case the decoupling
theorem remains valid, although there is the same mass dependence
as in the fermion case. As in this limit the bileptons become
almost degenerate, the custodial $\mathrm{SU_c(2)}$ symmetry
becomes almost exact. It is important to notice that the bilepton
gauge boson contribution to the $S$ parameter also vanishes in the
limit of exact degeneracy since $S\sim \log(\MX/\MY)$\cite{DNg3}.

Now let us go back to the static properties of the $W$ boson. It
turns out that a similar analysis as the one already presented can
be done for $\Delta Q$ and $\Delta \kappa$, though in this case
there is a more intricate mass dependence, which makes the
analysis less transparent [see Eqs.
(\ref{deltaQgen})-(\ref{f2keq})]. In the heavy mass limit we
have $\epsilon \ll \eta \sim \xi$, which yields Eqs.
(\ref{deltaQdeg}) and (\ref{deltakdeg}). In this case, shown in
Fig. \ref{Fig-decoupling}, both $\Delta Q$ and $\Delta \kappa$
vanish for a large bilepton mass: i.e. they are insensitive to a
heavy bilepton. In fact, from Eqs. (\ref{deltaQdeg}) and
(\ref{deltakdeg}) we get when $\epsilon \ll \eta \sim \xi$

\begin{equation}
\Delta Q\sim \Delta\kappa \sim
\frac{1}{\eta}=\left(\frac{m_W}{M_Y}\right)^2,
\end{equation}

\noindent which manifestly decouples from low-energy physics.

\begin{figure}
\centerline{\epsfig{file=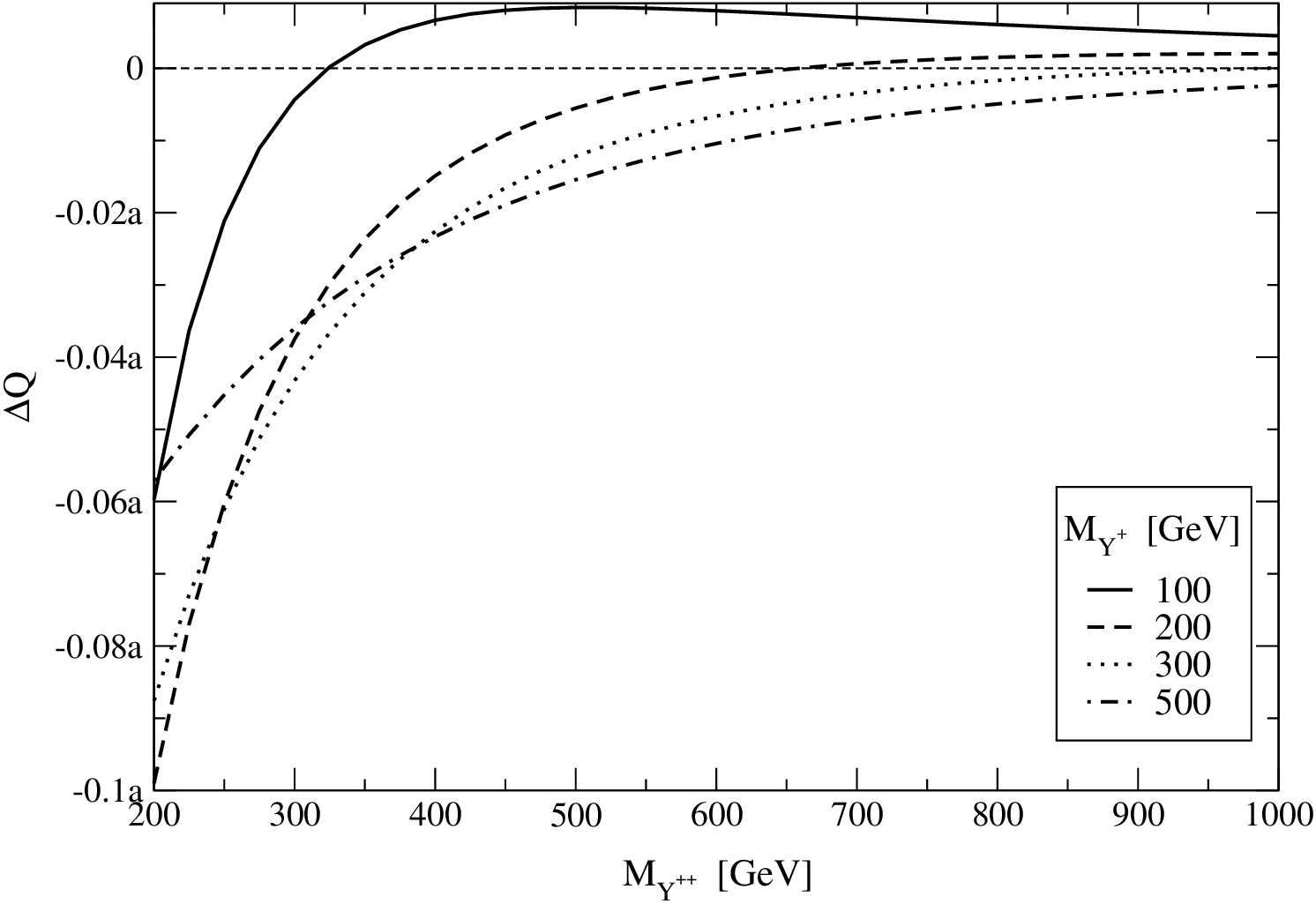,width=3in,clip=}}
\caption{Bilepton gauge boson contribution to the anomalous
$\Delta Q$ parameter, in units of $a=g^2/(96\,\pi^2)$, as a
function of the doubly charged gauge boson mass. We show curves
for diverse values of the singly charged gauge boson mass, as
indicated in the plot.} \label{Fig-deltaQc}
\end{figure}

\begin{figure}
\centerline{\epsfig{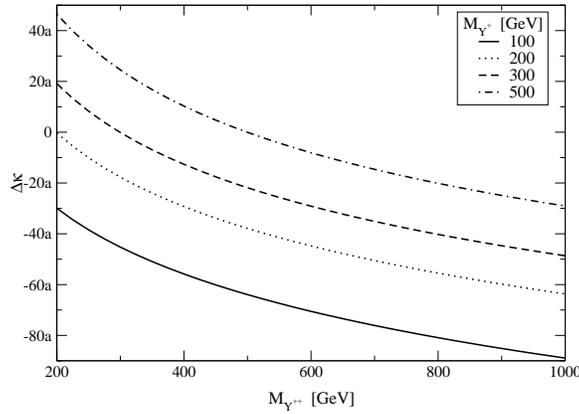}}
\caption{Bilepton gauge boson contribution to the anomalous
$\Delta \kappa$ parameter, in units of $a=g^2/(96\,\pi^2)$, as a
function of the doubly charged gauge boson mass. We show curves
for diverse values of the singly charged gauge boson mass, as
indicated in the plot.} \label{Fig-deltakc}
\end{figure}

We would like now to explore the hypothetical situation in which a
large mass splitting is allowed. It turns out that if we make
$\xi$ large while $\eta$ is kept fixed, $\Delta Q$ vanishes,
whereas $\Delta \kappa$ tends to a constant value. This scenario
is depicted in Figs. \ref{Fig-deltaQc} and \ref{Fig-deltakc}. In
Fig. \ref{Fig-deltaQc}, $\Delta Q$ is shown as a function of the
doubly charged bilepton mass, for diverse values of the singly
charged bilepton mass. It is evident that $\Delta Q$ would
decouple if $\MY$ would become heavy while $\MX$ remains fixed. On
the other hand, Fig. \ref{Fig-deltakc} shows a similar plot for
$\Delta \kappa$, which makes also evident that this parameter would be
sensitive to nondecoupling effects if the doubly charged bilepton
mass would become much heavier than the singly charged bilepton
mass. Although the situation illustrated in Figs.
\ref{Fig-deltaQc} and \ref{Fig-deltakc} is unrealistic within the
331 model, which forbids a mass splitting larger than the
electroweak scale, the previous analysis is useful to clarify the
following point: even in the scenario in which $\Delta \kappa$ is
sensitive to nondecoupling effects of a heavy particle, $\Delta Q$
is insensitive to such effects. This fact was noted in Ref.
\cite{Inami}, where it was explicitly verified that the
contributions to $\Delta Q$ from an extra fermion doublet and
technihadrons as well do decouple in the heavy mass limit
\cite{Inami}.

Finally, we would like to stress that the main difference with a
SM-like fermion doublet is that both components of the bilepton
doublet of the 331 model get a heavy mass from a large VEV ($u$),
which is heavier than the electroweak scale. On the other hand,
the splitting between the bilepton masses lies in the electroweak
scale since it arises from  VEVs which break the \sm~ gauge group
down to $\mathrm{U_e(1)}$. In fact, these VEVs  also give masses
to the SM gauge bosons. In the case of the SM-like fermion
doublet, its components acquire their masses from Yukawa
couplings. In summary, in the case of the bileptons, a heavy mass
implies a large VEV but not a large coupling, whereas in the
fermion case a large mass does implies a large coupling.  The
bilepton case has a close resemblance with the one discussed in
Ref. \cite{Li}, concerning a scalar doublet which acquires mass
from a bare parameter.

\section{Final remarks}
\label{remarks}

We have presented a detailed study of the bilepton gauge boson
contributions to the static properties of the $W$ boson. We have
presented explicit expressions for $\Delta Q$ and $\Delta \kappa$
in terms of elementary functions. We found that both $\Delta Q$
and $\Delta \kappa$ are of the same order of magnitude as those
contributions from other weakly coupled renormalizable theories,
like supersymmetric theories and the two-Higgs doublet model. An
important consequence of this result is that, unless an
overoptimistic precision is achieved in the future measurements of
the anomalous moments of the $W$ boson, it would be extremely
difficult to unravel the source of any possible deviation from the
SM, if such a deviation is detected indeed and arises from a
weakly coupled renormalizable theory. In the course of the last
Section, particular emphasis was given to the decoupling
properties of the bilepton gauge bosons. We have found that the
bilepton contribution to the static properties of the $W$ boson
decouples from low-energy physics as both the singly and the
doubly charged gauge boson masses become heavy. There is a
hypothetical scenario which might give rise to nondecoupling
effects, but it is unrealistic as involves a large mass splitting
(larger than the electroweak scale), which is not allowed in the
331 model since such a splitting is induced by the electroweak
scale. In this context, the bileptonic contribution has a close
resemblance with the contribution from a SM-like fermion doublet
or a scalar doublet. In fact, the contribution from some Feynman
diagrams involving bileptons has the same mass dependence as that
derived from the Feynman diagrams involving fermions or scalar
bosons. The main difference is that a large fermion mass comes
from a large Yukawa coupling, while a large bilepton mass requires
a large VEV. It has been argued that the last case does not give
rise to nondecoupling effects.

Finally, as a by-product of our calculation, we have studied  the
Yang-Mills sector which induces the interactions between the
bileptons and the SM gauge bosons. The respective trilinear and
quartic vertices have been studied and the Feynman rules were
derived within a nonlinear $R_\xi$ gauge covariant under the
$\mathrm{U_e(1)}$ gauge group, which allowed us to remove any
$\gamma YG_Y$ vertix. We hope that our results could be useful for
anyone interested in performing calculations involving these
couplings.

\acknowledgments{We acknowledge support from CONACYT and SNI (M\' exico).}

\appendix
\section{Couplings between the bileptons and the SM gauge bosons in the
331 model} \label{AppendixI}

In this Appendix we present explicit expressions for the vertices
arising from ${\cal L}_{{\mathrm{SMNP}}}$, which contains the
interactions between the SM gauge bosons and those predicted by
the 331 model.

\subsection{Trilinear vertices}

\begin{align}
{\cal L}_{WYY}&=\frac{i\,g}{\sqrt{2}}\bigg
[W^{+\mu}\left(Y^{--}_{\mu \nu}Y^{+\nu}-Y^+_{\mu
\nu}Y^{--\nu}\right)-W^+_{\mu
\nu}Y^{--\mu}Y^{+\nu}\nonumber \\
&-W^{-\mu}\left(Y^{++}_{\mu \nu}Y^{-\nu}-Y^-_{\mu
\nu}Y^{++\nu}\right)+W^-_{\mu \nu}Y^{++\mu}Y^{-\nu}\bigg ],
\end{align}
\begin{align}
{\cal L}_{\gamma YY}&=i\,e\bigg\{A^\mu \left(Y^-_{\mu
\nu}Y^{+\nu}-Y^+_{\mu \nu}Y^{-\nu}\right)-F_{\mu
\nu}Y^{-\mu}Y^{+\nu}\nonumber \\
&+2\left[A^\mu \left(Y^{--}_{\mu \nu}Y^{++\nu}-Y^{++}_{\mu
\nu}Y^{--\nu}\right)-F_{\mu \nu}Y^{--\mu}Y^{++\nu}\right]\bigg\},
\end{align}
\begin{align}
{\cal L}_{ZYY}&=\frac{ig}{2c_W}\bigg\{-\left(1+2\,s^2_W\right)\left[Z^\mu
\left(Y^-_{\mu \nu}Y^{+\nu}-Y^+_{\mu \nu}Y^{-\nu}\right)-Z_{\mu
\nu}Y^{-\mu}Y^{+\nu}\right]\nonumber \\
&+ \left(1-4\,s^2_W\right)\left[Z^\mu (Y^{--}_{\mu
\nu}Y^{++\nu}-Y^{++}_{\mu \nu}Y^{--\nu})-Z_{\mu
\nu}Y^{--\mu}Y^{++\nu}\right]\bigg\}.
\end{align}

\subsection{Quartic vertices}

\begin{align}
{\cal L}_{WWYY}&=-\frac{g^2}{2}\Big \{W^{+\mu}\left[
Y^{+\nu}\left(W^-_\mu Y^-_\nu-W^-_\nu
Y^-_\mu\right)+Y^{--\nu}\left(W^-_\mu Y^{++}_\nu-W^-_\nu
Y^{++}_\mu\right)\right]\nonumber \\
&+\left(W^+_\mu W^-_\nu-W^-_\mu
W^+_\nu\right)\left(Y^{--\mu}Y^{++\nu}-Y^{-\mu}Y^{+\nu}\right)\Big
\},
\end{align}

\begin{align}
{\cal L}_{\gamma
WYY}&=-\frac{ge}{\sqrt{2}}\left(Q_{Y^+}+Q_{Y^{++}}\right)A^\mu \big
[Y^{--\nu}\left(W^+_\mu Y^+_\nu-W^+_\nu Y^+_\mu\right)\nonumber \\
&+Y^{++\nu}\left(W^-_\mu
Y^-_\nu-W^-_\nu Y^-_\mu\right)\big ],
\end{align}

\begin{align}
{\cal L}_{ZWYY}&=\frac{-g^2}{2\,\sqrt{2}\,c_W} Z^\mu\Big
\{\left(1-4\,s^2_W\right)\left[Y^{++\nu}\left(W^-_\mu
Y^-_\nu-W^-_\nu Y^-_\mu \right)+Y^{--\nu}\left(W^+_\mu
Y^+_\nu-W^+_\nu
Y^+_\mu \right)\right] \nonumber \\
&-\left(1+2\,s^2_W\right)\left[Y^{+\nu}\left(W^+_\mu
Y^{--}_\nu-W^+_\nu Y^{--}_\mu \right)+Y^{-\nu}\right(W^-_\mu
Y^{++}_\nu-W^-_\nu Y^{++}_\mu\left)\right] \nonumber \\
&+2\,c^2_W\left[W^{+\nu}\left(Y^{--}_\mu Y^+_\nu-Y^{--}_\nu
Y^+_\mu \right)+W^{-\nu}\left(Y^{++}_\mu Y^-_\nu-Y^{++}_\nu
Y^-_\mu \right)\right]\Big \},
\end{align}

\begin{equation}
{\cal L}_{\gamma \gamma YY}=-e^2\,A^\mu \left[Y^{+\nu}\left(A_\mu
Y^-_\nu-A_\nu Y^-_\mu\right)+4\,Y^{++\nu}\left(A_\mu
Y^{--}_\nu-A_\nu Y^{--}_\mu\right)\right],
\end{equation}

\begin{align}
{\cal L}_{\gamma ZYY}&=-\frac{g\,e}{2\,c_W}Z^\mu \Big
\{-\left(1+2\,s^2_W\right)\left[Y^{+\nu}\left(A_\mu Y^-_\nu-A_\nu
Y^-_\mu\right)+Y^{-\nu}\left(A_\mu Y^+_\nu-A_\nu Y^+_\mu\right)\right]\nonumber \\
&+2\left(1-4\,s^2_W\right)\left[Y^{++\nu}\left(A_\mu
Y^{--}_\nu-A_\nu Y^{--}_\mu\right)+Y^{--\nu}\left(A_\mu
Y^{++}_\nu-A_\nu Y^{++}_\mu\right)\right]\Big \},
\end{align}

\begin{align}
{\cal L}_{ZZYY}=&-\frac{g^2}{4\,c^2_W}Z^\mu \bigg[
\left(1+2\,s^2_W\right)^2Y^{+\nu}\left(Z_\mu Y^-_\nu-Z_\nu
Y^-_\mu\right)\nonumber \\
&+ \left(1-4s^2_W\right)^2Y^{++\nu}\left(Z_\mu Y^{--}_\nu-Z_\nu
Y^{--}_\mu\right)\bigg].
\end{align}

\noindent In the above expressions, $V_{\mu \nu}=\partial_\mu
V_\nu-\partial_\nu V_\mu$ ($V=\gamma,\,Z,\, W,\, Y$). We have
omitted those vertices which arise from the last term of Eq.
(\ref{LSMNP}) because they involve the neutral ${Z^\prime}$ boson.

\section{Feynman rules in a $\mathrm{U_{\lowercase{e}}(1)}$-covariant gauge}
\label{AppendixII}

This gauge is defined by means of the following nonlinear
gauge-fixing functions, which transform covariantly under the
$\mathrm{U_e\left(1\right)}$ group:

\begin{align}
f^{++}_Y&=D^e_\mu Y^{++\mu}-i\xi\, \MY\, G^{++}_Y, \\
f^+_Y&=D^e_\mu Y^{+\mu}-i\,\xi\, \MX\, G^+_Y,
\end{align}

\noindent where $D^e_\mu=\partial_\mu-i\,e\,Q_Y\,A_\mu$
($Q_Y=1,\,2$) is the $\mathrm{U_e(1)}$ covariant derivative, $\xi$
is the gauge parameter, and $G_Y$ are the pseudo-Goldstone bosons
associated with the bilepton gauge fields.

This gauge allows us to eliminate the $YG_Y\gamma$ vertices but not
the $YWG_Y$ ones, which are given by

\begin{equation} {\cal
L}_{YWG}=-\frac{i\,g}{\sqrt{2}}\left[W^{+\mu}\left(G^+_Y\partial_\mu
G^{--}_Y-G^{--}_Y\partial_\mu
G^+_Y\right)-W^{-\mu}\left(G^-_Y\partial_\mu
G^{++}_Y-G^{++}_Y\partial_\mu G^-_Y\right)\right].
\end{equation}

The interactions between the pseudo-Goldstone bosons and the
photon obey scalar electrodynamics:

\begin{equation}
{\cal L}_{G_YG_Y\gamma}=\left(D^e_\mu G^+_Y\right)^\dag
\left(D^{e\,\mu}G^+_Y\right)+ \left(D^e_\mu G^{++}_Y\right)^\dag
\left(D^{e\,\mu}G^{++}_Y\right).
\end{equation}

The gauge-fixing Lagrangian can be written in the form

\begin{align}
{\cal L}_{{\mathrm
GF}}&=-\frac{1}{\xi}f^{-}_Yf^+_Y-\frac{1}{\xi}f^{--}_Yf^{++}_Y
\nonumber \\
&=-\frac{1}{\xi}\left(D^e_\mu Y^{+\mu}\right)^\dag \left(D^e_\nu Y^{+\nu}\right)
-\frac{1}{\xi}\left(D^e_\mu Y^{++\mu}\right)^\dag \left(D^e_\nu Y^{++\nu}\right)-\xi
\SMX G^-_YG^+_Y\nonumber \\
&-\xi \SMY G^{--}_YG^{++}_Y +i\,\MX\left(G^+_Y\left(D^e_\mu
Y^{+\mu}\right)^\dag-G^-_Y\left(D^e_\nu
Y^{+\nu}\right)\right)\nonumber \\
&+i\,\MY\left(G^{++}_Y\left(D^e_\mu
Y^{++\mu}\right)^\dag-G^{--}_Y\left(D^e_\nu
Y^{++\nu}\right)\right).
\end{align}

\noindent After integration by parts, the last two terms of this expression
cancel out the bilinear, $YG_Y$, and the trilinear, $YG_Y\gamma$, couplings that
arise from the Higgs kinetic-energy sector.

Finally, the Faddeev-Popov Lagrangian needed for the calculation
of the $WW\gamma$ vertex has the following form

\begin{align}
{\cal L}_{{\mathrm
FPG}}=&\left(D^{e\,\mu\dag}\bar{C}_Y^{--}\right)\left(D^e_\mu
C^{++}_Y\right)+
\left(D^{e\,\mu}\bar{C}_Y^{++}\right)\left(D^{e\dag}_\mu
C^{--}_Y\right)-\xi
\SMY\left(\bar{C}^{--}_YC^{++}_Y+\bar{C}^{++}_YC^{--}_Y\right)\nonumber\\
&+ \left(D^{e\,\mu\dag}\bar{C}_Y^-\right)\left(D^e_\mu
C^+_Y\right)+
\left(D^{e\,\mu}\bar{C}_Y^+\right)\left(D^{e\dag}_\mu
C^-_Y\right)-\xi
\SMX\left(\bar{C}^-_YC^+_Y+\bar{C}^+_YC^-_Y\right) \nonumber \\
&+\frac{i\,g}{\sqrt{2}}\bigg\{W^{+\mu}\left[\left(D^{e\dag}_\mu
\bar{C}^{--}_Y\right)C^+_Y-\left(D^e_\mu
\bar{C}^+_Y\right)C^{--}_Y\right]\nonumber \\
&- W^{-\mu}\left[\left(D^e_\mu
\bar{C}^{++}_Y\right)C^-_Y-\left(D^{e\dag}_\mu
\bar{C}^-_Y\right)C^{++}_Y\right]\bigg\}
\end{align}

\begin{figure}
\centerline{\epsfig{file=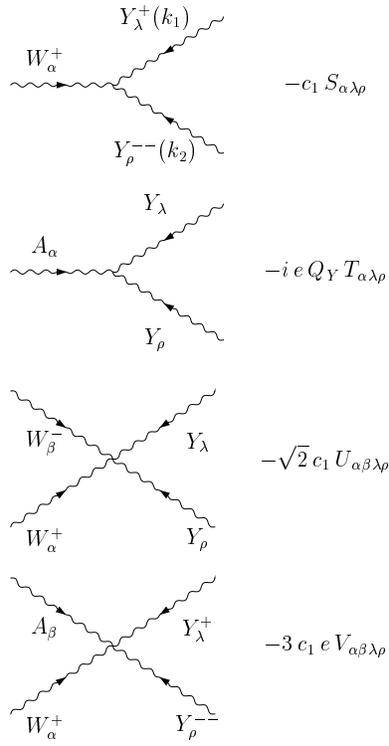,width=2in,clip=}}
\caption{Feynman rules for the trilinear and quartic vertices
involving singly and doubly charged gauge bosons.
$c_1=i\,g/\sqrt{2}$ and the respective expressions for $S$, $T$,
$U$ and $V$ are given in Table \ref{table-gauge}.}
\label{Fig-bgfr}
\end{figure}

\begin{figure}
\centerline{\epsfig{file=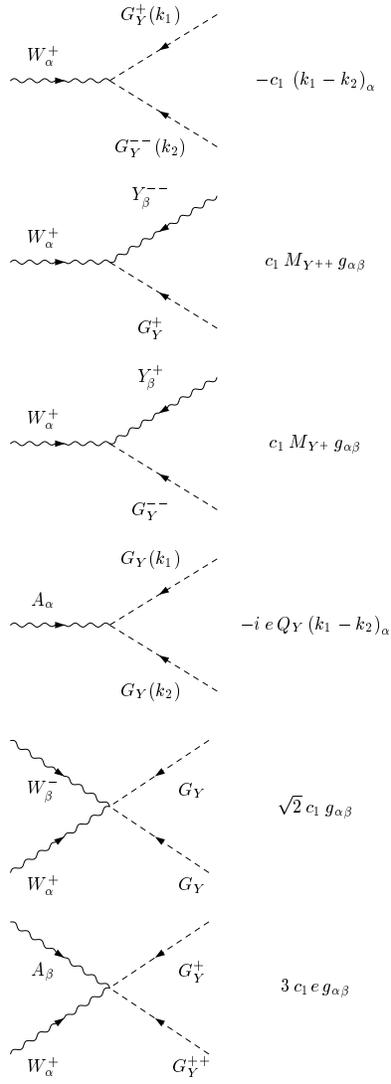,width=2in,clip=}}
\caption{Feynman rules for the trilinear and quartic vertices
involving singly and doubly charged pseudo-Goldstone bosons.}
\label{Fig-gbfr}
\end{figure}

\begin{figure}
\centerline{\epsfig{file=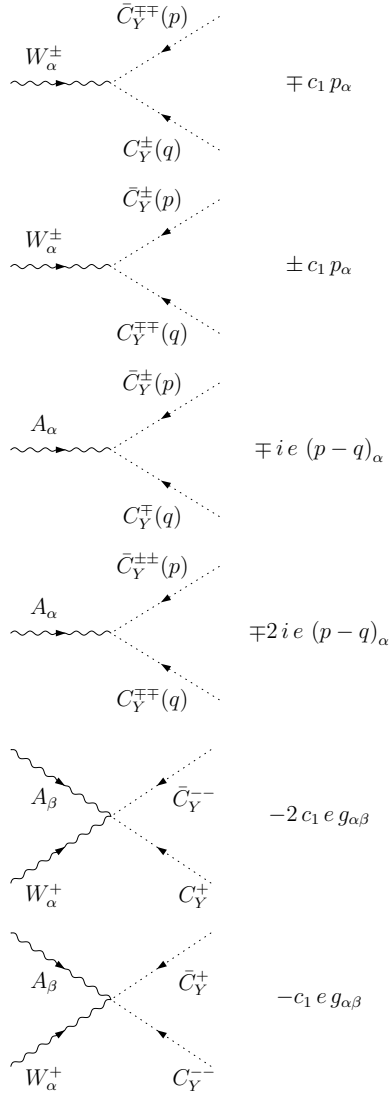,width=2in,clip=}}
\caption{Feynman rules for the trilinear and quartic vertices
involving singly and doubly charged ghost fields.}
\label{Fig-ghfr}
\end{figure}

The respective Feynman rules are summarized in Figs.
\ref{Fig-bgfr}-\ref{Fig-ghfr} and Table \ref{table-gauge}. It can
be seen that QED-like Ward identities are satisfied by the
$YY\gamma$, $G_YG_Y\gamma$, and $\bar{C}_YC_Y\gamma$ vertices.

\begin{table}
\caption{Feynman rules for the vertices shown in Fig.
\ref{Fig-bgfr}. $U$ and $U^\prime$ stand for the $WWYY$ vertex
when $Y=Y^{+}$ and $Y=Y^{++}$, respectively.}
\begin{tabular}{cc}
&$\mathrm{U_e(1)}$-covariant gauge\\
\hline
$S_{\alpha\,\lambda\,\rho}$&$(k_2-k_1)_\alpha g_{\lambda
\rho}+(k-k_2)_\lambda g_{\alpha \rho}+(k_1-k)_\rho g_{\alpha
\lambda}$\\
$T_{\alpha\,\lambda\,\rho}$&$(k_2-k_1)_\alpha g_{\lambda \rho}+
(k-\frac{1}{\xi}\,k_1-k_2)_\lambda g_{\alpha
\rho}+(k_1+\frac{1}{\xi}\,k_2-k)_\rho g_{\alpha \lambda}$\\
$U_{\alpha\,\beta\,\lambda\,\rho}$&$g_{\alpha\,\beta}\,g_{\lambda\,\rho}
+g_{\alpha\,\rho}\,g_{\beta\,\lambda}-2\,g_{\alpha\,\lambda}\,g_{\beta\,\rho}$\\
$U^\prime_{\alpha\,\beta\,\lambda\,\rho}$&$g_{\alpha\,\beta}\,g_{\lambda\,\rho}
-2\,g_{\alpha\,\rho}\,g_{\beta\,\lambda}+g_{\alpha\,\lambda}\,g_{\beta\,\rho}$\\
$V_{\alpha\,\beta\,\lambda\,\rho}$&$g_{\alpha\,\beta}\,g_{\lambda\,\rho}
-\,g_{\alpha\,\rho}\,g_{\beta\,\lambda}$\\
\end{tabular}
\label{table-gauge}
\end{table}


\begin{references}


\bibitem{Bardeen}W. A. Bardeen, R. Gastmans, and B. Lautrup, Nucl. Phys. {\bf
B46}, 319 (1972); See also E. N. Argyres, {\it et al.}, Nucl.
Phys. {\bf B391}, 23 (1993).

\bibitem{Couture1} G. Couture and J. N. Ng, Z Phys. C {\bf 35}, 65
(1987).

\bibitem{Couture2} G. Couture, J. N. Ng, J. L. Hewett, and T. G. Rizzo,
Phys. Rev. D {\bf 36}, 859 (1987).

\bibitem{SUSY} C. L. Bilachak, R. Gastmans, and A. van Proeyen, Nucl. Phys. {\bf
B273}, 46 (1986); G. Couture, J. N. Ng, J. L. Hewett, and T. G.
Rizzo, Phys. Rev. D {\bf 38}, 860 (1988); A. B. Lahanas and V. C.
Spanos, Phys. Lett. B {\bf 334}, 378 (1994); T. M. Aliyev, Phys.
Lett B {\bf 155}, 364 (1985).

\bibitem{Sharma} N. K. Sharma, P. Saxena, Sardar Singh, A. K. Nagawat,
and R. S. Sahu, Phys. Rev. D {\bf 56}, 4152 (1997).

\bibitem{Rizzo} T. G. Rizzo and M. A. Samuel, Phys. Rev. D {\bf 35},
403 (1987); A. J. Davies, G. C. Joshi, and R. R. Volkas, Phys.
Rev. D {\bf 42}, 3226 (1990).

\bibitem{Larios} F. Larios, J. A. Leyva, and R. Mart\'\i nez, Phys.
Rev. D {\bf 53}, 6686 (1996).

\bibitem{Wudka} For a review on the $WW\gamma$ coupling within the
effective Lagrangian approach, see J. Ellison and J. Wudka, Annu.
Rev. Nucl. Part. Sci. {\bf 48}, 33 (1998); and references therein.

\bibitem{Pleitez1}F. Pisano and V. Pleitez, Phys. Rev. {\bf D 46}, 410
(1992); P. H. Frampton, Phys. Rev. Lett. {\bf 69}, 2889 (1992).

\bibitem{Toscano}See for instance G. Tavares-Velasco
and J. J. Toscano, Europhys. Lett. {\bf 53}, 465 (2001).

\bibitem{Appelquist} T. Appelquist and J. Carazzone, Phys. Rev. D {\bf
11}, 2856 (1975).

\bibitem{nondecoupling}See for instance J. Collins, F. Wilczek, and A. Zee,
Phys. Rev. D {\bf 18}, 242 (1978); D. Toussaint, Phys. Rev. D {\bf
18}, 1626 (1978); L. H. Chan, T. Hagiwara, and B. Ovrut, Phys.
Rev. D {\bf 20}, 1982 (1979).

\bibitem{Li} L. F. Li, Z. Phys. C {\bf 58}, 519 (1993).

\bibitem{Inami} T. Inami, C. S. Lim, B. Takeuchi, and M.
Tanabashi, Phys. Lett. B {\bf 381}, 458 (1996).

\bibitem{DNg1} D. Ng, Phys. Rev. D {\bf 49}, 4805 (1994).

\bibitem{Veltman1} G.  Passarino and M. Veltman, Nucl. Phys. {\bf B160}, 151
(1979); see also A. Denner, Fortschr. Phys. {\bf 41}, 307 (1993).

\bibitem{Stuart}R. G. Stuart, Comp. Phys. Commun. {\bf 48}, 367 (1988);
see also G. Devaraj and R.G. Stuart, Nucl. Phys. {\bf B519}, 483
(2000).

\bibitem{DNg2} J. T. Liu and D. Ng, Phys. Rev. D {\bf 50}, 548
(1994).

\bibitem{Long} H. N. Long and D. Van Soa, Nucl. Phys. {\bf B601}, 361 (2001).

\bibitem{DNg3} J. T. Liu and D. Ng, Z. Phys. C {\bf 62}, 693
(1994).

\bibitem{Fujikawa} K. Fujikawa, Phys. Rev. D {\bf 7}, 393 (1973); M.
Ba\' ce and N. D. Hari Dass, Ann. Phys. (N. Y.) {\bf 94}, 349
(1975); M. B. Gavela, G. Girardi, C. Malleville, and P. Sorba,
Nucl. Phys. {\bf B193}, 175 (1981); U. Cotti, J. L. D\'\i az-Cruz,
and J. J. Toscano, Phys. Lett. B {\bf 404}, 308 (1997); Phys. Rev.
D {\bf 62}, 035009 (1990); J. M. Hern\' andez, M. A. P\' erez, G.
Tavares-Velasco, and J. J. Toscano, Phys. Rev. D {\bf 60}, 013004
(1990).

\bibitem{Mertig} R. Mertig, M. B\"{o}hm and A.  Denner, Comp. Phys. Commun.
{\bf 64}, 345 (1991).

\bibitem{Veltman2} G. t' Hooft and M. Veltman, Nucl. Phys. {\bf B153}, 365
(1979).

\bibitem{FF} G. J. van Oldenborgh, Comput. Phys. Commun. {\bf 66} 1 (1991); T.
Hahn and M. P\'erez-Victoria hep-ph/9807565.

\bibitem{bileptons}F. Cuypers and S. Davidson, Eur. Phys. J. C
{\bf 2}, 503 (1998); B. Dion, {\it et al.}, Phys. Rev. D {\bf 59},
075006 (1999);

\bibitem{Frampton}P. H. Frampton, Int. J. Mod. Phys. A {\bf 13}, 2345 (1998).

\bibitem{Willman} L. Willmann {\it et al.}, Phys. Rev. Lett. {\bf 82}, 49
(1999).

\bibitem{Pleitez2} V. Pleitez, Phys. Rev. D {\bf 61} 057903 (2000); P. Das and P. Jais,
Phys. Rev. D {\bf 62}, 075001 (2000); P. H. Frampton and A. Rasin,
Phys. Lett B. {\bf 482}, 129 (2000).

\bibitem{Joshi} M. B. Tully and G. C. Joshi, Phys. Lett. B {\bf 466}, 333 (1999).

\bibitem{Gregores}E. M. Gregores, A. Gusso, and S. F. Novaes, Phys. Rev. D {\bf 64},
015004 (2001).

\bibitem{Joshi2}M. B. Tully and C. G. Joshi, Int. J. Mod. Phys A {\bf 13}, 5593
(1998).

\bibitem{Veltman3}M. Veltman, Nucl. Phys. {\bf B123}, 89 (1977).

\bibitem{Li2} T. P. Cheng and L. F. Li, Phys. Rev. D {\bf 44}, 1502 (1991).


\end{references}
\end{document}